\documentclass{svjour3}                     
\smartqed  
\usepackage{graphicx,natbib,amsmath}
\usepackage{amsmath,amsfonts,amssymb,bm,url,color}

\newcommand{\EQ}{\begin{equation}}
\newcommand{\EN}{\end{equation}}
\newcommand{\eq}[1]{(\ref{#1})}

\newcommand{\Eq}[1]{Eq.~(\ref{#1})}
\newcommand{\Eqs}[2]{Eqs.~(\ref{#1}) and~(\ref{#2})}

\newcommand{\Eqps}[2]{Eqs.~(\ref{#1}) and (\ref{#2})}
\newcommand{\eqss}[2]{(\ref{#1})--(\ref{#2})}

\newcommand{\Fig}[1]{Figure~\ref{#1}}

\newcommand{\Tab}[1]{Table~\ref{#1}}

\newcommand{\Figs}[2]{Figs.~\ref{#1} and \ref{#2}}

\def\half{{\textstyle{1\over2}}}

\newcommand{\qq}{\bm{q}}
\newcommand{\bra}[1]{\langle #1\rangle}
\newcommand{\ysci}[5]{: #1, #5, {\em Science }{\bf #2}, #3--#4.}
\newcommand{\ynat}[5]{: #1, #5, {\em Nature }{\bf #2}, #3--#4.}
\newcommand{\ypr}[5]{: #1, #5, {\em Phys.\ Rev. }{\bf #2}, #3--#4.}
\newcommand{\yprl}[5]{: #1, #5, {\em Phys.\ Rev.\ Lett. }{\bf #2}, #3--#4.}
\newcommand{\yprlN}[4]{: #1, #4, {\em Phys.\ Rev.\ Lett. }{\bf #2}, #3.}

\newcommand{\ypreN}[4]{: #1, #4, {\em Phys.\ Rev.\ E }{\bf #2}, #3.}
\newcommand{\yrrpN}[4]{: #1, #4, {\em Rep.\ Prog.\ Phys. }{\bf #2}, #3.}

\newcommand{\yab}[5]{: #1, #5, {\em Astrobiol. }{\bf #2}, #3--#4.}
\newcommand{\ypnas}[5]{: #1, #5, {\em Proc.\ Nat. Acad.\ Soc. }{\bf #2}, #3--#4.}
\newcommand{\yoleb}[5]{: #1, #5, {\em Orig.\ Life Evol.\ Biosph. }{\bf #2}, #3--#4.}
\newcommand{\yijaS}[5]{: #1, #5 {\em Int.\ J.\ Astrobiol. }{\bf #2}, #3--#4.}
\newcommand{\yjcp}[5]{: #1, #5, {\em J.\ Chem.\ Phys. }{\bf #2}, #3--#4.}
\newcommand{\yjcpN}[4]{: #1, #4, {\em J.\ Chem.\ Phys. }{\bf #2}, #3.}
\newcommand{\yjour}[6]{: #1, #6, {\em #2} {\bf #3}, #4--#5.}
\newcommand{\yjourN}[5]{: #1, #5, {\em #2} {\bf #3}, #4.}
\newcommand{\ybook}[3]{: #1, {\em #2}, #3.}

\begin{document}

\title{The limited roles of autocatalysis and enantiomeric cross inhibition in achieving homochirality in dilute systems}
\titlerunning{Homochirality in dilute systems}

\author{Axel Brandenburg
}

\institute{
Nordita, KTH Royal Institute of Technology and Stockholm University, 
Department of Astronomy, AlbaNova University Center,
Stockholm University, SE-10691 Stockholm, Sweden;\\
JILA and Laboratory for Atmospheric and Space Physics,
Boulder, CO 80303, USA\\
\\
Correspondence:
Nordita, Roslagstullsbacken 23, SE-10691 Stockholm, Sweden\\
\email{brandenb@nordita.org}, Tel: +46 8 5537 8707, mobile: +46 73 270 4376\\
\url{http://orcid.org/0000-0002-7304-021X}
}

\date{\today}

\maketitle


\begin{abstract} 
To understand the effects of fluctuations on achieving homochirality,
we employ a Monte-Carlo method where autocatalysis and enantiomeric
cross-inhibition, as well as racemization and deracemization reactions
are included.
The results of earlier work either without autocatalysis or without
cross-inhibition are reproduced.
Bifurcation diagrams and the dependencies of the number of reaction
steps on parameters are studied.
In systems with 30,000 molecules, for example, up to a billion reaction
steps may be needed to achieve homochirality without autocatalysis.
\end{abstract}


\section{Introduction}

There are many reasons why the chemistry of life is based on carbon.
The ability to form complex macromolecules is one of them
\citep{RGS08,Lon14}.
Many carbon-bearing molecules also have the property of being chiral,
i.e., the three-dimensional structure of such a molecule is different
from its mirror image \citep{AGK91}.
Even relatively simple amino acids such as alanine, arginine, and proline
are such examples.
In solution, these molecules rotate polarized light to the left, which is
why they are called levorotatory, so we say they are of the {\sc l} form.
Only rarely does life on Earth involve amino acids of the {\sc d} form
\citep{MWR05}.
Those molecules are dextrorotatory and would rotate polarized light in
the opposite direction.
Naturally occurring sugars such as ribose in the backbone of DNA and
RNA are also chiral and of the {\sc d} form.

Abiogenically produced carbon-bearing chiral molecules always occur as
a mixture of {\sc d} and {\sc l} forms.
One calls such mixtures racemic, i.e., the molecules are achiral
as opposed to chiral, i.e., the mixture is not homochiral and
molecules of both the {\sc d} and {\sc l} forms occur.
However, synthesizing chiral polymers such as proteins and nucleic acids
only succeeds in the predominately isotactic form.
This realization goes back to the experiments of \cite{Joyce84} using
template-directed polycondensation.
This work demonstrated what was called enantiomeric cross-inhibition,
i.e., the termination of polycondensation with enantiomers of the
opposite chirality.
More recent work, however, showed that enantiomeric cross-inhibition
may not always operate \citep{Joyce14} and may also not
be very efficient on the early Earth.
This was the reason for \cite{JBG15,JBG17} to question the long-held
belief in the necessity of enantiomeric cross-inhibition for producing
homochirality based on the famous proposal of \cite{Frank}.
In that work, \cite{Frank} suggested that the interplay of autocatalysis
combined with what he called mutual antagonism would be a necessary
ingredient for reaching complete homochirality.
\cite{JBG15,JBG17} invoked the possibility that local fluctuations
could yield results that are different from those expected by solving
the kinetic rate equations.
Fluctuation effects are generally expected to play a role in small
compartments, or when concentrations are low.

In an earlier paper, \cite{SHS08} did already invoke the effects of
fluctuations and questioned the concept of autocatalysis.
Indeed, the number of known autocatalytic reactions is small -- with
the reaction of \cite{Soai} being basically still the only one.
However, the more general case of arbitrary combinations with and without
autocatalysis and with and without enantiomeric cross-inhibition has
not yet been studied.
Such a more general approach is of interest in view of numerical studies
such as that of \cite{Tox13}, where homochirality has been found without
apparent autocatalysis or enantiomeric cross-inhibition.

Both the model of \cite{Frank} that is based on rate equations and the
alternative stochastic models discussed above describe what is known as
spontaneous chiral symmetry breaking.
These mechanisms do not require an external chiral influence, even though
a very small chiral influence is always present.
Possible candidates for causing a systematic chiral influence
include polarized light from the UV radiation
in star-forming regions \citep{Bailey98,Bailey01} or from nearby
neutron stars \citep{Bon99} or the weak force in fundamental physics
\citep{HRS80,Hegstrom,MT84}.
The expected effect is likely to be very small \citep{Bada95,Len06}.
This can be quantified by studying the interplay between spontaneous
chiral symmetry breaking and the effect of an external chiral influence
\citep{KN83,KN85}.
We expect those conclusions to carry over to the stochastic models
as well.
To verify this, we allow in some of our models for the presence of a
chiral influence in the catalytic effect, as was done in the context of
a polycondensation model using non-stochastic rate equations \citep{BAHN}.
They found that, as the fidelity of the autocatalytic reactions is
increased, a typical bifurcation diagram emerges.
Owing to the effect of an external chiral influence, the diagram becomes
slightly asymmetric with respect to positive and negative enantiomeric
excess.
It is therefore referred to as an imperfect bifurcation, just as it was
anticipated by \cite{KN83,KN85}.

\section{Method}

The notion of invoking both autocatalysis and mutual antagonism is based
on the governing rate equations for the three reactions
\begin{align}
A+D &\stackrel{k_{\rm C}~}{\longrightarrow} 2D,\label{AD}\\
A+L &\stackrel{k_{\rm C}~}{\longrightarrow} 2L,\label{AL}\\
D+L &\stackrel{k_\times~}{\longrightarrow} 2A,\label{DL}
\end{align}
where $A$ denotes an achiral molecule that can autocatalyze at a rate
$k_{\rm C}$ to a chiral one either of the {\sc d} or the {\sc l} form,
while $D$ and $L$ can cross-inhibit into an achiral one at a rate
$k_\times$.
However, rate equations no longer provide a suitable description of
the relevant kinetics when the system is dilute and reactions are rare
\citep{Gil77,Tox14}.
In that case, a stochastic approach must be adopted.
Developing an intuitive and simple Monte Carlo method will be the goal
of the present paper.

It will be advantageous to regard the reaction rates as {\em probabilities}
for the respective reactions to occur within a given reaction step.
Thus, instead of solving the underlying master equation, as was done by
\cite{SHS08,SHS09} and \cite{JBG15,JBG17}, we adopt here a Monte-Carlo
approach in which at each reaction step $n$, one of several possible
reactions occur \citep{Ver67}.
The state of the system is governed by the state vector
\begin{equation}
\qq_n=([A],\,[D],\,[L]),
\end{equation}
where squared brackets denote the concentrations of the respective
molecules, and $n$ denotes the reaction step.
At each reaction step $n$, we select a transition $\Delta \qq_n$ out
of a set of seven possible transitions such that the new state vector
$\qq_{n+1}$ is given by
\begin{equation}
\qq_{n+1}=\qq_n+\Delta \qq_n.
\end{equation}
For example, for the three reactions \eqss{AD}{DL}, $\Delta\qq_n$
could be one of the three vectors $(-1,\,1,\,0)$, $(-1,\,0,\,1)$,
or $(2,\,-1,\,-1)$.

Let us illustrate the formalism using an example.
Initially, at $n=0$, the system has, say, 10 molecules of each
of the three possible ones ($A$, $D$, and $L$), so we have
$\qq_0=(10,\,10,\,10)$.
Suppose now that reaction \eq{AD} takes place during the next reaction
step, then $\Delta \qq_0=(-1,\,1,\,0)$, so in our example we would have
in the next step $\qq_1=(9,\,11,\,10)$.
This means that one out of the 10 molecules of the {\sc d} form reacted
with an achiral one $A$ to produce two new molecules of the {\sc d} form.
One of the {\sc A} molecules got removed, so only 9 of them are left.
Also one of the $D$ molecules got removed, but since one of them did
already participate in the reaction, only one more of them has occurred
after the reaction, i.e., we now have 11 molecules of the {\sc d} form.
Note that the model is mass conserving, i.e., the total number of
molecules is constant and always equal to the initial value $N$.
No polycondensation reactions are considered here.

In addition to the reactions discussed above, we also allow for
spontaneous racemization and deracemization reactions, i.e.,
\begin{align}
&A \stackrel{k_+~}{\longrightarrow} D, \quad & 
 A \stackrel{k_+~}{\longrightarrow} L, \label{ADAL} \\
&D \stackrel{k_-~}{\longrightarrow} A, \quad &
 L \stackrel{k_-~}{\longrightarrow} A. \label{DLA}
\end{align}
Spontaneous racemization can be the result of natural degradation;
see, e.g., \cite{Bada+70}.
Spontaneous deracemization was introduced by \cite{SHS08}.
Note that the $\Delta \qq_n$ of the pair of reactions in \Eqps{AD}{AL}
is the same as in \Eq{ADAL}.
The former two reactions, which are autocatalytic, can also be written
in a form similar to \Eqps{ADAL}{DLA} by replacing $k_+$ by $k_{\rm C}[D]/N$
for the first reaction and $k_{\rm C}[L]/N$ for the second one.
Here, $N=[D]+[L]+[A]$ is the total number of all molecules, regardless
of whether they are chiral ($D$ or $L$) or achiral ($A$).
Likewise, the reaction \eq{DL} corresponds to racemization reactions
\eq{ADAL} with the coefficients $k_-$ in the two parts of that equation
being replaced by $k_\times[L]/N$ and $k_\times[D]/N$, respectively.
Thus, for the autocatalytic and enantiomeric cross-inhibition reactions,
we have
\begin{align}
&A \stackrel{\underrightarrow{~k_{\rm C}[D]/N~}}{} D,\label{ADkC}\\
&A \stackrel{\underrightarrow{~k_{\rm C}[L]/N~}}{} L,\label{ALkC}
\end{align}
\begin{equation}
\quad\begin{cases}
D \stackrel{\underrightarrow{~k_\times([D]+[L])/N~}}{} A,\\
L \stackrel{\underrightarrow{~k_\times([D]+[L])/N~}}{} A.
\end{cases}
\label{DALA}
\end{equation}
The curly bracket to the left on the last two reactions indicates
that those two reactions happen at the same time.

To study the effects of an imperfect fidelity and of an external chiral
influence on the system, we extend our model in a way that is analogous
to the formulation employed by \cite{BAHN}.
This is accomplished by changing \Eqs{ADkC}{ALkC} to
\begin{align}
&A \stackrel{\underrightarrow{~k_{\rm C}(f_+[D]+f_-[L]+\beta_D[A])/N~}}{} D,\label{ADkC2}\\
&A \stackrel{\underrightarrow{~k_{\rm C}(f_+[L]+f_-[D]+\beta_L[A])/N~}}{} L,\label{ALkC2}
\end{align}
where $f_\pm=(1\pm f)/2$ with $0\leq f\leq1$ being the fidelity ($f=1$
for perfect fidelity) and $\beta_{D/L}$ quantifies the bias of the system
toward $D$ or $L$, respectively ($\beta_D=\beta_L=0$ in the absence of
an external chiral influence on the system).

We recall that in this paper we regard the coefficients
$k_+$, $k_-$, $k_{\rm C}$, and $k_\times$ not as rates,
but as probabilities.
The sum of all probabilities must then not exceed unity, i.e.,
the four coefficients must obey the constraint
\begin{equation}
k_+ +k_- +k_{\rm C}+k_\times\leq1.
\label{MassConstraint}
\end{equation}
We summarize all the reactions modelled in this paper in \Tab{Tab}.

\begin{table}[b!]\caption{Summary of all seven reactions (6)--(10).
}\vspace{12pt}\centerline{\begin{tabular}{clcccccc}
$i$ & $w_i$ & $\Delta\qq$ & reaction & equation \\
\hline
1 & $\half k_+$        & $(-1,\,1,\,0)$ & $A\stackrel{k_+~}{\longrightarrow}D$ & \eq{ADAL} \\
2 & $\half k_+$        & $(-1,\,0,\,1)$ & $A\stackrel{k_+~}{\longrightarrow}L$ & \eq{ADAL} \\
3 & $\half k_-$        & $(1,\,-1,\,0)$ & $D\stackrel{k_-~}{\longrightarrow}A$ & \eq{DLA} \\
4 & $\half k_-$        & $(1,\,0,\,-1)$ & $L\stackrel{k_-~}{\longrightarrow}A$ & \eq{DLA} \\
5 & $k_{\rm C}\,[D]/N$ & $(-1,\,1,\,0)$ & $A+D\stackrel{k_{\rm C}~}{\longrightarrow}2D$ & \eq{ADkC} \\
6 & $k_{\rm C}\,[L]/N$ & $(-1,\,0,\,1)$ & $A+L\stackrel{k_{\rm C}~}{\longrightarrow}2L$ & \eq{ALkC} \\
7 & $k_\times([D]+[L])/N$ & $(2,\,-1,\,-1)$ & $D+L\stackrel{k_\times~}{\longrightarrow}2A$ & \eq{DALA} \\
\label{Tab}\end{tabular}}\end{table}

The coefficients $k_+$, $k_-$, $k_{\rm C}$, and $k_\times$ are
used to determine seven intervals, $w_1$, $w_2$, ..., $w_7$, with
$w_1+w_2+...+w_7=1$.
The values of $w_i$ are equal to the respective probabilities
of the seven possible reactions \eqss{ADAL}{DALA}.
The two reactions in \eq{ADAL} occur with equal probability,
so we set $w_1=w_2=k_+/2$.
Likewise, the two reactions in \eq{DALA} occur with equal probability,
so we set $w_3=w_4=k_-/2$.
Finally, we have $w_5=k_{\rm C}[D]/N$, $w_6=k_{\rm C}[L]/N$, and
$w_7=k_\times([D]+[L])/N$.
Since $([D]+[L])/N$ is, in general, less than unity, it may be
possible that no reaction occurs during a particular step.
This is true even if $\beta_D\neq0$ and $\beta_L\neq0$.
(Both $\beta_D$ and $\beta_L$ are below unity.)

We denote the interval boundaries by $r_i$, so we have
$0\leq r_1\leq r_2\leq ... \leq r_7\equiv1$.
At reach reaction step, we generate a new random number $r$ with
$0\leq r\leq1$, where the value of $r$ determines which of the
various reactions occurs.
Reaction $i$ is performed when
\begin{equation}
r_{i-1} \leq r < r_i\quad\mbox{(for reaction $i$ with $i=1$, $2$, ..., $7$)}.
\end{equation}
The widths of the seven intervals, $r_i-r_{i-1}=w_i$, are listed in \Tab{Tab}.

To make statistically meaningful statements, we have to perform
sufficiently many experiments, and then consider averages over all
experiments.
In practice, we perform many experiments at the same time and refer
to them as populations.
The populations are completely independent of each other.
We then compute normalized averages over all populations, denoted by
angle brackets, so $\bra{A}$, $\bra{D}$, and $\bra{L}$ are the
fractional concentrations of each of the three types of molecules.
Therefore, we have
\begin{equation}
\bra{A}+\bra{D}+\bra{L}=1.
\end{equation}
Within each population, the enantiomeric excess is defined as
\begin{equation}
\eta=\frac{[D]-[L]}{[D]+[L]}.
\end{equation}
It can be between $-1$ and $+1$ for homochiral states and is close
to zero for a racemic mixture.
From a statistical point of view, however, the two homochiral states,
$\eta=\pm1$, are equivalent, so only the average of the modulus of
$\eta$ is of primary interest.
To determine for which parameters the bifurcation occurs, we monitor the
mean enantiomeric excess $\bra{|\eta|}$, which is close to zero if there
are about equally many molecules of the {\sc d} and of the {\sc l} forms.

Altogether, our models have five parameters:
$k_+$, $k_-$, $k_{\rm C}$, $k_\times$,
and the population size $N$.
The number of populations is an additional parameter that is
chosen to be 512 in all cases.
The models are completely symmetric with respect to $D$ and $L$,
i.e., there is no preference with respect to $D$ and $L$.

\begin{figure}[t!]\begin{center}
\includegraphics[width=.49\textwidth]{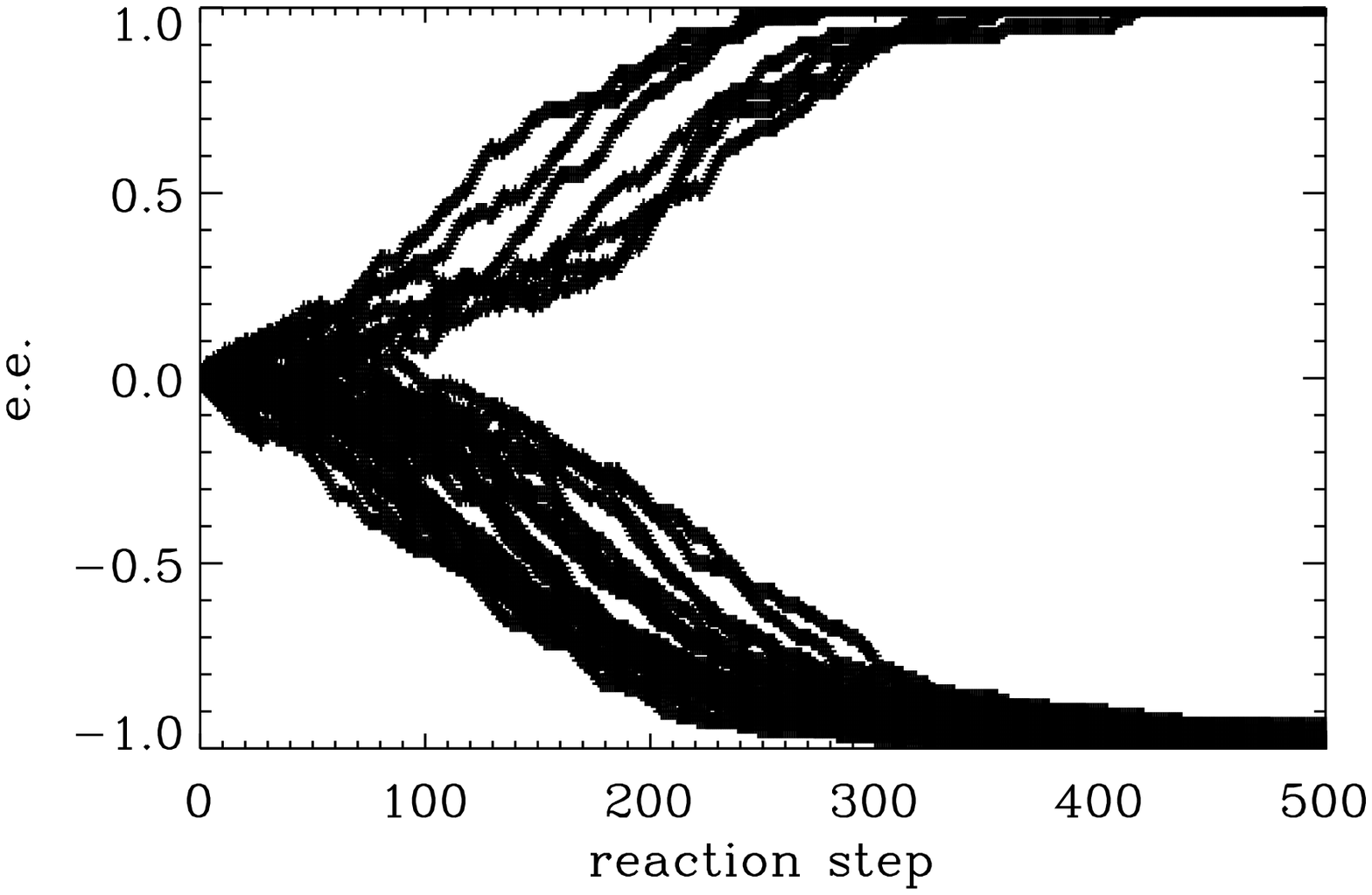}
\includegraphics[width=.49\textwidth]{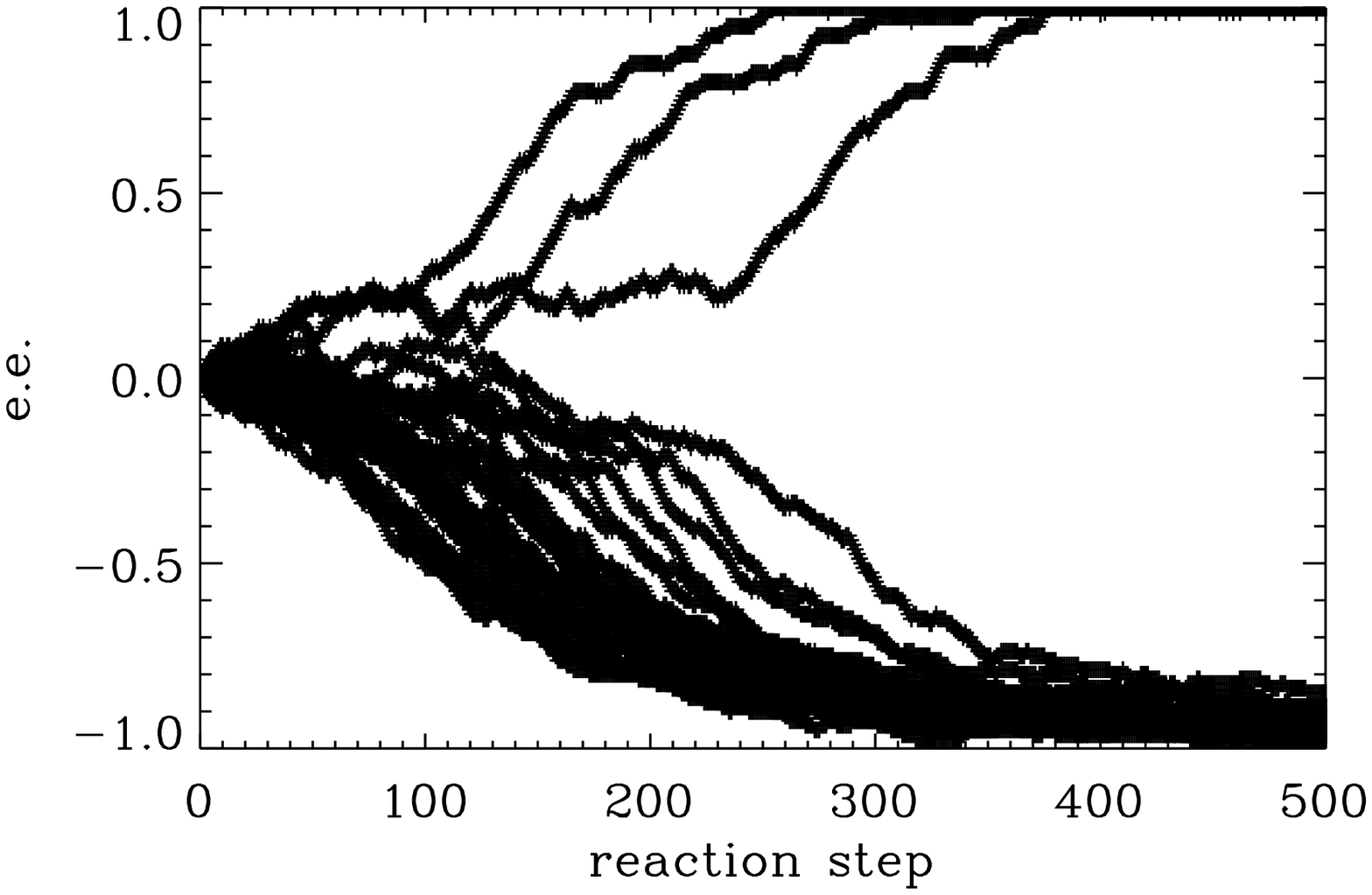}
\includegraphics[width=.49\textwidth]{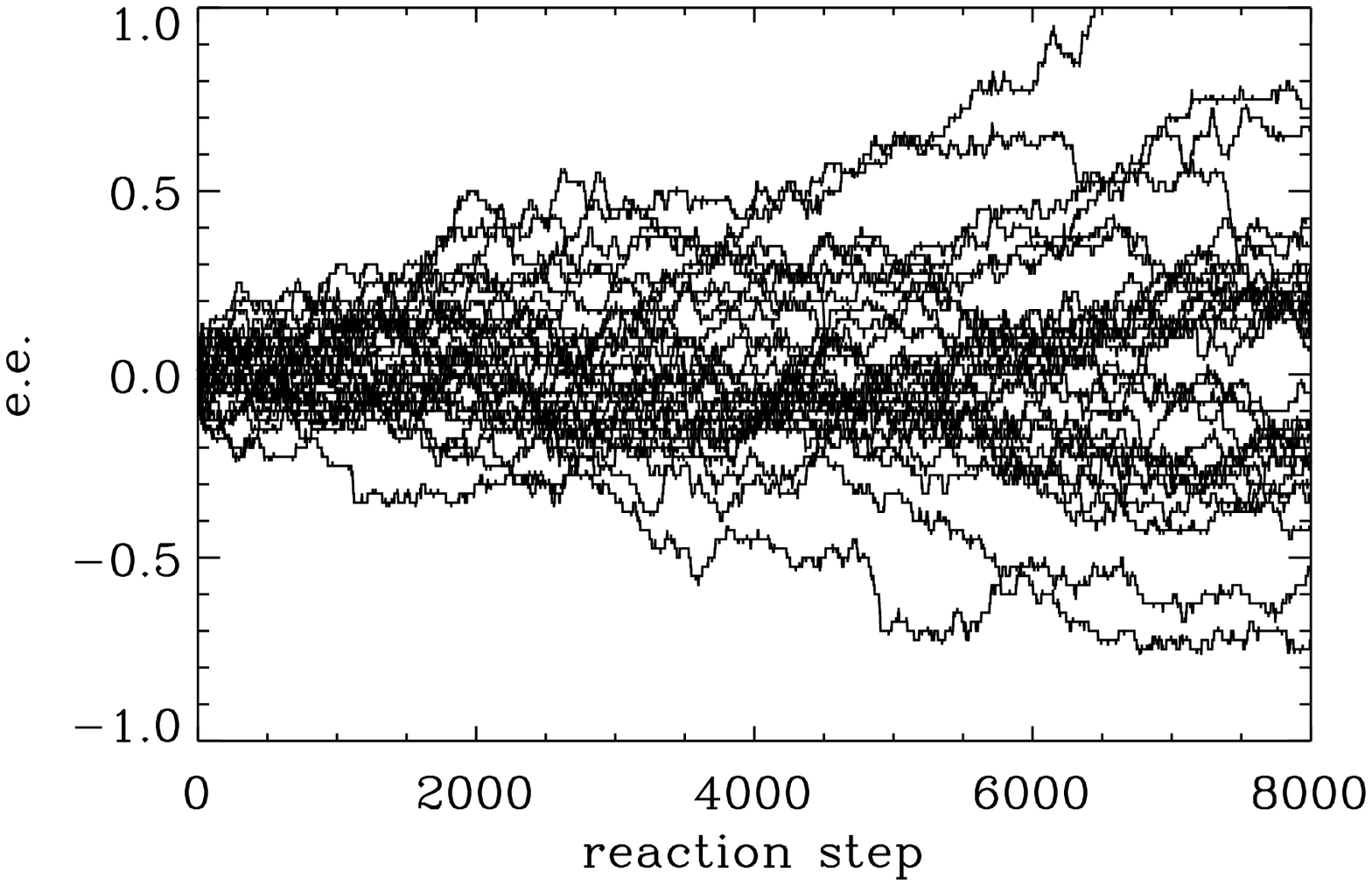}
\includegraphics[width=.49\textwidth]{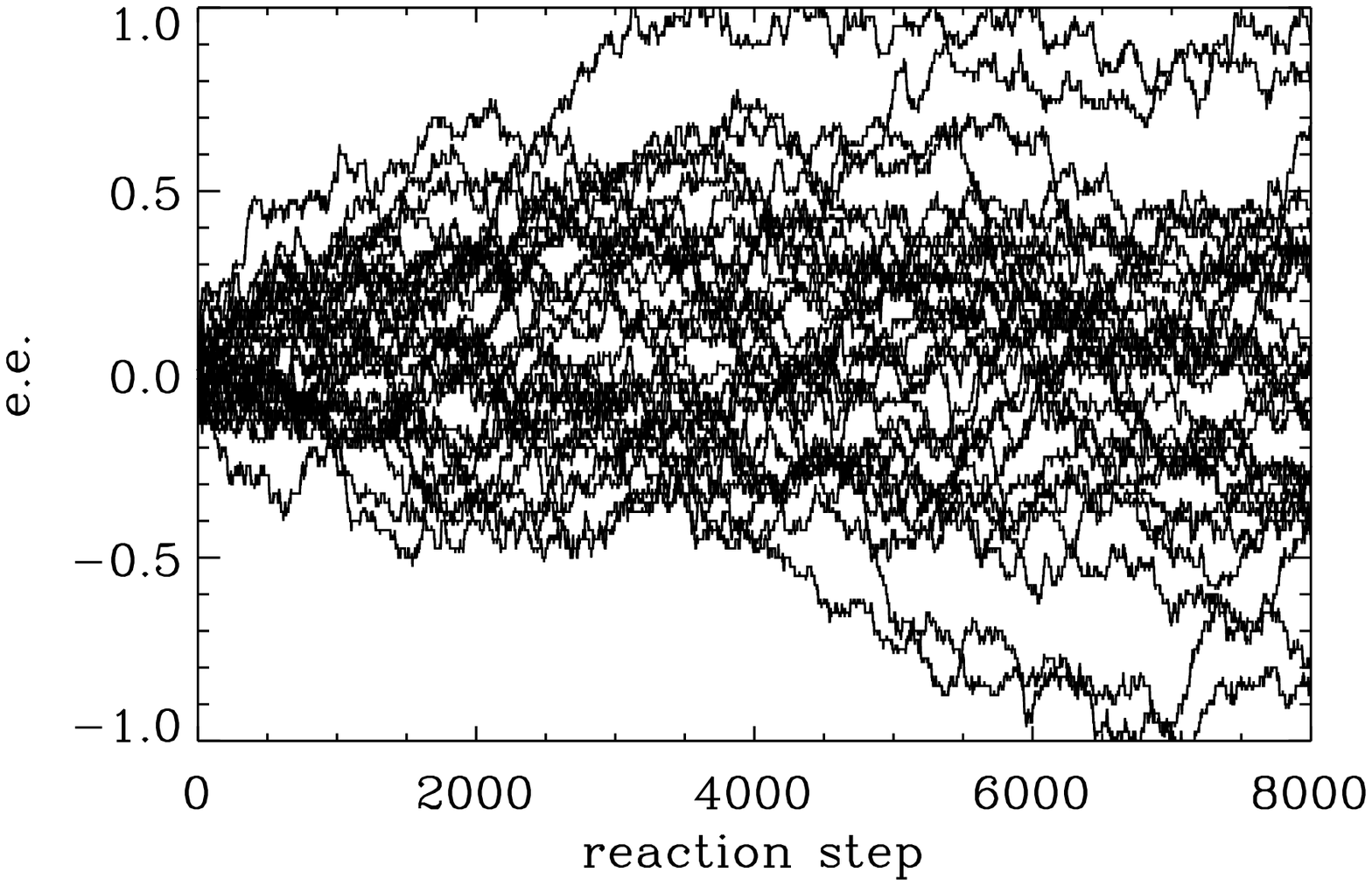}
\end{center}\caption[]{
Examples of the evolution of the enantiomeric excess (e.e.)
for cases I, II, III, and IV.
{\em Upper left}: model~I for $k_{\rm C}=1-k_\times=0.1$;
{\em upper right}: model~II for $k_{\rm C}=1-k_-=0.9$;
{\em lower left}: model~III for $k_+=1-k_\times=0.3$;
{\em lower right}: model~II for $k_+=1-k_-=0.3$.
}\label{pwave}\end{figure}

We report here the results of numerical experiments performed with
the {\sc Pencil Code}\footnote{\url{https://github.com/pencil-code},
DOI:10.5281/zenodo.2315093}, a publicly available time stepping code that
is designed to perform computations on massively parallel computers.
It uses the third order Runge--Kutta time stepping scheme of \cite{Wil80}.
Normally, the code is used for solving partial differential equations
using meshpoints.
Here, however, no spatial extent will be considered and each mesh point
can be regarded as an independent population.
This is how many populations can then be solved for in parallel.
The time step is chosen to be unity to reproduce a discrete reaction step.
Furthermore, the derivative module {\tt noderiv} is in the code, which
means that no extra ghost zones are used, which would only be needed
in a model with spatial extent.

\section{Results}

\subsection{Strategy}

We begin by studying the familiar case of autocatalysis and enantiomeric
cross-inhibition, i.e., we choose the value of $k_\times$ and, since
$k_+=k_-=0$ in this case, we set $k_{\rm C}=1-k_\times$, so the bound
given by \Eq{MassConstraint} is saturated.
We refer to this as model~I, which corresponds to that of \cite{Frank}.
As already explained above, each reaction step can correspond to a
different time interval, which does not need to be specified for our
present purpose.
Next, we consider the case proposed by \cite{JBG15,JBG17}, where we set
$k_+=k_\times=0$ and vary $k_{\rm C}$ such that $k_{\rm C}+k_-=1$.
This is referred to as model~II.
The case proposed by \cite{SHS08} corresponds to $k_-=k_{\rm C}=0$,
so we vary $k_+$ and adjust $k_\times=1-k_+$.
This is model~III.
Finally, we consider the case $k_{\rm C}=k_\times=0$, vary $k_+$,
and adjust $k_-=1-k_+$.
This is our model~IV.
We also consider intermediate cases that we refer to as I/III and
II/IV, where we have considered {\em three} non-vanishing probabilities,
and a model V, where all four probabilities are non-vanishing;
see \Tab{Tab2} for an overview of all the seven models.
For models I, II, III, and IV, we vary $p$ with $0<p<1$, while for
models I/III and II/IV, we vary $q$, but now in the range $0<q<p$,
keeping $p=0.4$ fixed.
When $q=0$, model~I/III is identical to model~I, while for $q=p$,
model~I/III is identical to model~III.
Likewise, model~II/IV is identical to model~II for $q=0$
and identical to model~IV for $q=p$.
For model~V, we only consider one case where $k_+=k_-=k_\times=0.2$
and $k_{\rm C}=0.4$.

\begin{table}[htb]\caption{
Summary of the sets of experiments discussed in this paper.
The columns ``auto'' and ``inhib'' indicate where autocatalysis
and enantiomeric cross-inhibition are possible.
}\vspace{12pt}\centerline{\begin{tabular}{clcccccl}
Model & $k_+$ & $k_-$ & $k_{\rm C}$ & $k_\times$ & auto & inhib & ref.\\
\hline
I    &  0  &   0   &  $p$  & $1-p$ & yes & yes & \cite{Frank}\\
II   &  0  & $1-p$ &  $p$  &   0   & yes & no & \cite{JBG15} \\
III  & $p$ &   0   &   0   & $1-p$ & no  & yes & \cite{SHS08} \\
IV   & $p$ & $1-p$ &   0   &   0   & no     & no  &  --- \\
I/III& $q$ &   0   & $p-q$ & $1-p$ & partly & yes & ---  \\
II/IV& $q$ & $1-p$ & $p-q$ &   0   & partly & no  & --- \\
V    & 0.2 &  0.2  &  0.4  &  0.2  & partly & yes & --- \\
\label{Tab2}\end{tabular}}\end{table}

\begin{figure}[t!]\begin{center}
\includegraphics[width=\textwidth]{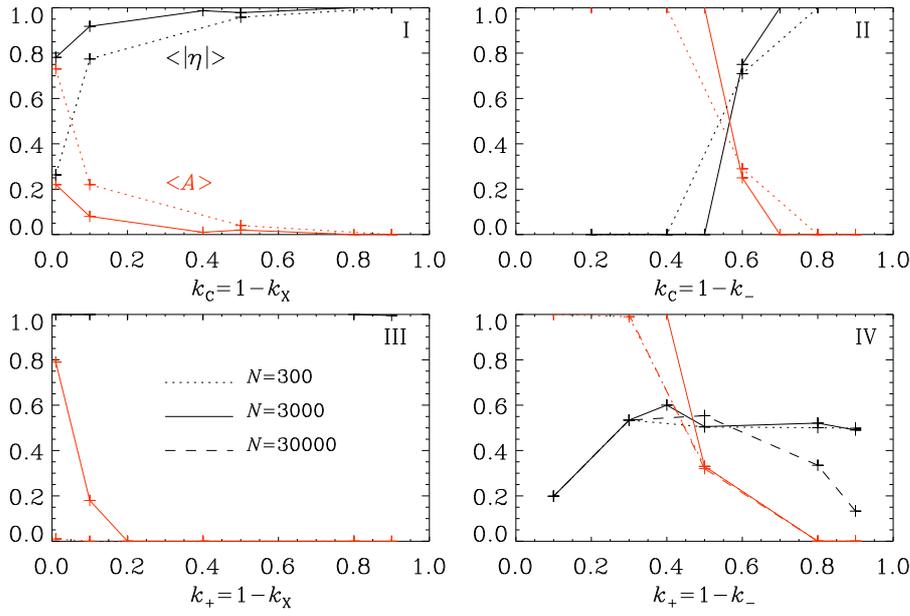}
\end{center}\caption[]{
Bifurcation diagrams of $\bra{|\eta|}$ (black) and $\bra{A}$ (red)
for $N=3000$ (solid lines) and $N=300$ (dotted lines)
as a function of parameters for models~I, II, III, and IV.
}\label{pmodels2}\end{figure}

\subsection{Models~I, II, III, and IV}

In models~I, II, and III, there is a bifurcation of solutions within each
population to either $+1$ or $-1$, depending on chance.
To determine for which parameters a bifurcation occurs, we monitor
the mean unsigned enantiomeric excess, $\bra{|\eta|}$.
It will be close to unity if most of the achiral $A$ molecules
of the substrate are turned into $D$ or $L$.
Thus, we expect $\bra{A}$ two vary inversely with $\bra{|\eta|}$.
This is indeed the case.
For model~IV, however, we see that $\bra{A}$ gradually changes from
$1$ for $k_+\leq0.2$ to $\bra{A}=0$ for $k_+\geq0.8$.
Thus, for large values of $k_+$, there are hardly any achiral molecules
left, but this does not mean that the final stage is chiral.
Instead, there are large fluctuations among different populations,
where some of the molecules can be close to fully chiral, with many
of them being of the {\sc d} form in some populations, and many
of them being of the {\sc l} form in other populations.
This is particularly the case for smaller populations with, e.g.,
300 or 3000 members.
For $30,000$ members, on the other hand, some strongly chiral states
are still possible when $k_+$ is not too large.
Indeed, we see that the black dashed line in \Fig{pmodels2} shows a
maximum for $k_+=1-k_-=0.5$.
However, this only happens after a significant number of reaction steps.

\begin{figure}[t!]\begin{center}
\includegraphics[width=\textwidth]{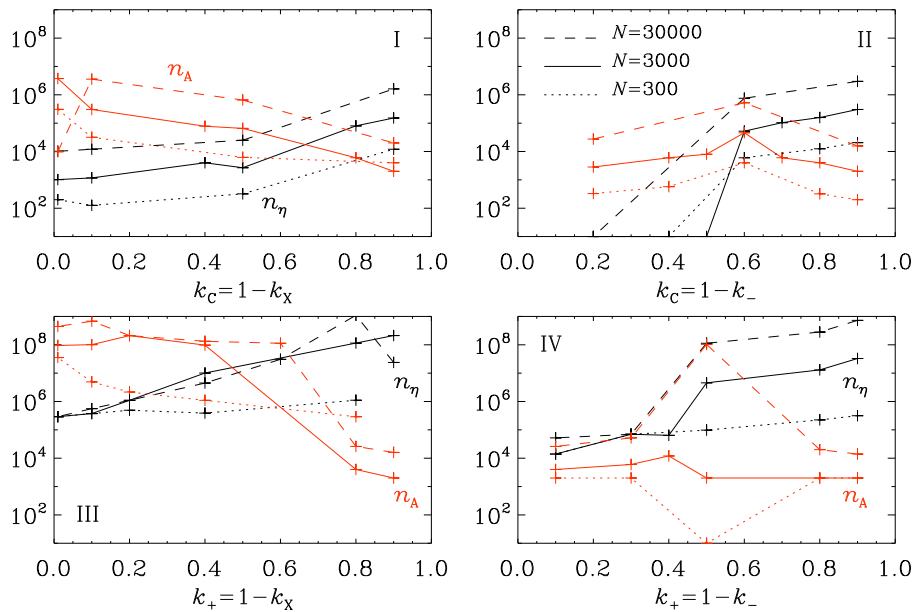}
\end{center}\caption[]{
Typical reaction numbers $n_\eta$ (black lines) and $n_A$ (red lines)
as a function of parameters for models~I, II, III, and IV for $N=3000$
(solid lines) and $N=300$ (dotted lines).
}\label{pmodels}\end{figure}

\begin{figure}[t!]\begin{center}
\includegraphics[width=\textwidth]{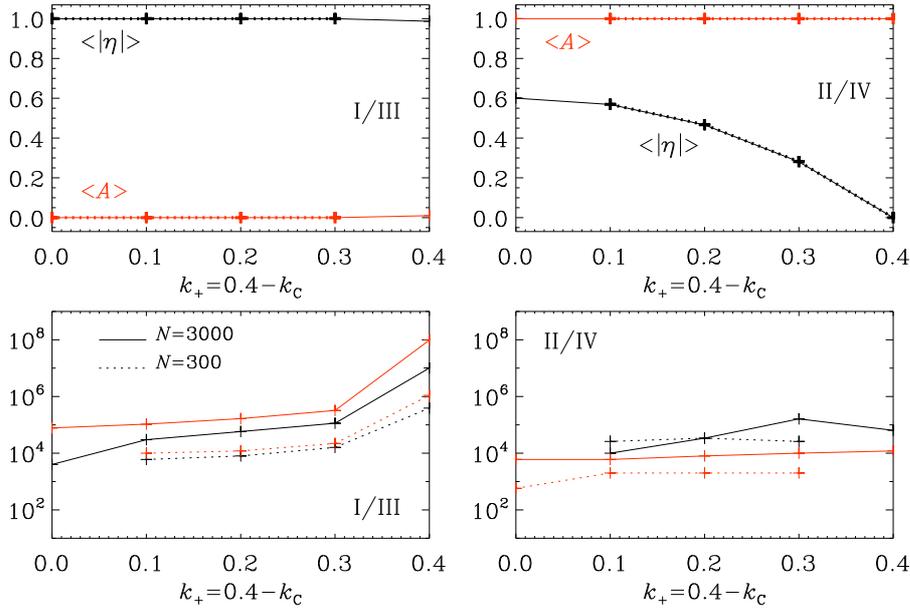}
\end{center}\caption[]{
Bifurcation diagrams of $\bra{|\eta|}$ (black) and $\bra{A}$ (red) (upper row)
and time scales $n_\eta$ (black lines) and $n_A$ (red lines) (lower row)
for $N=3000$ (solid lines) and $N=300$ (dotted lines) 
for the mixed cases~I/III (left, for $k_\times=0.6$)
and II/IV (right, for $k_-=0.6$) as a function of $k_+$,
keeping $k_{\rm C}=0.4-k_+$ in each case.
}\label{pmixed}\end{figure}

To characterize the necessary number of reaction steps for different
parameter combinations, we define the parameters $n_\eta$ and $n_A$ as
the reaction step after which $\bra{|\eta|}$ and $\bra{A}$, respectively,
have reached their final values within $1\%$.
In \Fig{pmodels}, we plot these numbers as a function of parameters for
different population sizes.

For models~III and IV, $n_\eta$ can be extremely large -- of the
order of $10^8$ -- especially when the population size is large.
For model~IV with $N=30,000$, for example, even $n_\eta=10^9$ is reached
when $k_+\to1$ and $k_-\to0$.
An increasing trend of $n_\eta$ with $k_+$ is also found for model~III,
i.e., the model of \cite{SHS08}, again for large population sizes.
In those cases, there is also a dramatic drop in $n_A$ when $k_+\to1$,
which shows that most of the molecules are either of the {\sc d} or of
the {\sc l} form.

For models~I and II, on the other hand, both $n_\eta$ and $n_A$ are
significantly smaller -- typically between $10^3$ and $10^6$.
Nevertheless, we see again an increasing trend of $n_\eta$ and a
decreasing trend of $n_A$ when $k_{\rm C}$ is increased, which is
analogous to the increase with $k_+$ in models~III and IV.

\subsection{Models~I/III and II/IV}

The models~I/III and II/IV were designed to assess the possibility of
cooperative effects between autocatalysis ($k_{\rm C}$) and spontaneous
deracemization ($k_+$).
In other words, can we trade some fraction of autocatalytic reactions
for deracemization and still have the same or an even stronger
effect on achieving homochirality?

\begin{figure}[t!]\begin{center}
\includegraphics[width=\textwidth]{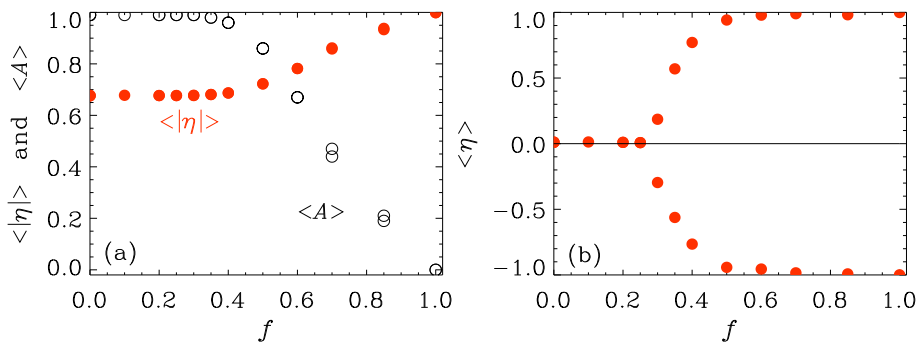}
\end{center}\caption[]{
(a) $\bra{|\eta|}$ (red) together with $\bra{A}$ (black) versus $f$
and (b) $\bra{\eta}$ versus $f$ for model~V with $N=300$,
$k_+=k_-=k_\times=0.2$, $k_{\rm C}=0.4$,
$\beta_{\rm D}=10^{-3}$, and $\beta_{\rm L}=0$.
Note the slight asymmetry for positive and negative values of $\bra{\eta}$.
}\label{ppsum}\end{figure}

Looking at \Fig{pmixed}, we see that for model~I/III, the values of
$\bra{|\eta|}$ and $\bra{A}$ are unchanged as $k_+=0.4-k_{\rm C}$ is
changed, but both $n_\eta$ and $n_A$ increase as $k_+$ is increased
and thus $k_{\rm C}$ decreased.
This suggests that there is no cooperative effect of spontaneous
deracemization on autocatalysis, because it now takes more steps
to achieve the same result.
For model~II/IV, we see that $\bra{A}$ is again unchanged, but now
$\bra{|\eta|}$ decreases.
Thus, replacing some of the autocatalytic reactions by deracemization
has  a detrimental effect.
The values of $n_\eta$ increase with increasing $k_+$, while $n_A$
remains of the order of $10^4$, except for some departures at $k_+=0.5$.

\subsection{Model~V with finite fidelity and an external chiral influence}

It may have appeared strange that in \Figs{pmodels2}{pmixed},
$\bra{|\eta|}$ was always close to unity -- even for rather
small values of $k_{\rm C}$.
The reason is that, until now, autocatalysis was modelled with perfect
fidelity ($f=1$), i.e., the presence of molecules of the {\sc d} form
always favors the production of more molecules of the {\sc d} form and
not of the {\sc l} form, and vice versa.
This is different when $f<1$ in \Eqs{ADkC2}{ALkC2}.

Models with $f<1$ were first studied by \cite{San03} in a fairly detailed
polycondensation model.
Later, \cite{BAHN} also allowed for the presence of a small bias
($\beta_D\neq0$ or $\beta_L\neq0$) in such a model.
In those cases, a bifurcation diagram emerges, where $\bra{\eta}\neq0$
for $f=0$.
This is referred to as an imperfect bifurcation, because it is asymmetric
with respect to positive and negative values of $\bra{\eta}$.

We employ what we call here model~V, where, in addition to autocatalysis
and enantiomeric cross-inhibition, also spontaneous racemization and
deracemization are included.
Adding these two effects has the advantage that they cause the system
to be well ``mixed''.
Without this, even the case of $f=0$ can in some cases lead to a gradual
loss of all achiral molecules, which is unrealistic.
This problem is alleviated by having $k_+\neq0$ and $k_-\neq0$.

The result for model~V is shown in \Fig{ppsum}, where we plot not only
$\bra{|\eta|}$, but now also $\bra{\eta}$, which is the average of the
signed enantiomeric excess over all populations.
Unlike $\bra{|\eta|}$, this quantity can approach $+1$ or $-1$ only
if a large number of independent populations or geneses produce the
same chirality.
We recall that the number of populations is still 512.
The number of members of each population is here taken to be 300.
This model shows an imperfect bifurcation at $f\approx0.2$.
There is a small neighborhood around this point where for small enough
perturbations only positive values of $\bra{\eta}$ are possible.

Realistic values of $\beta_{\rm D}$ or $\beta_{\rm L}$ could be in the
range $10^{-12}$ to $10^{-17}$; see \cite{KN85}, who demonstrated that
even such a small chiral influence good tip the balance systematically
in one direction.
This ignores, however, the effects of fluctuations associated with small
system sizes.
In practice, this would always dominate over thermal fluctuations.
It would therefore remain an exciting prospect to look for biomolecules
of unconventional chirality, for example, through metabolic experiments
on Mars, as proposed by \cite{Sun09}.

\section{Conclusions}

For many decades, the paradigm of \cite{Frank} of producing
homochirality by a combination of both autocatalysis and mutual
antagonism or enantiomeric cross-inhibition has shaped much of our
thinking for many decades.
It is now clear that the implied necessity of the combined presence
of both of these ingredients may not strictly hold.
Earlier numerical simulations of \cite{Tox13} have already hinted at
such a possibility, which may well be a viable one in small enough
systems where fluctuations can be important.
Two separate aspects of this were already studied in some detail
by \cite{SHS08,SHS09} and \cite{JBG15,JBG17}.

In this work, we have presented a unified approach to homochirality
by considering all possible combinations discussed above: with and without
autocatalysis, and with and without enantiomeric cross-inhibition.
In most of the cases, we have allowed for population sizes between
30 and 30,000 members.
Our approach allows us to understand the earlier work of
\cite{SHS08,SHS09} and \cite{JBG15,JBG17} in the broader context of
models that include these two models as special cases.
In fact, the close relation between these approaches does not seem
to have been broadly recognized yet.
The only paper that mentions both of them is the recent review of
\cite{Wal17}.

The unified approach to stochastic effects in chemical systems
discussed in the present paper is conceptually simple and can
easily be generalized to other systems, for example those
with additional spatial extent.
Such an approach is particularly important to astrobiology and the origin
of life, given that independent geneses may have occurred at different
locations on the early Earth \citep{DL05} and that concentrations may
have been low.
Although some of the processes reported above may involve altogether
up to a billion reaction steps, this may not be long on geochemical time
scales and would correspond to only about 40,000 years if we assumed a
reaction time of 20 minutes.
Independent geneses may yield opposite chiralities at different
locations on the early Earth \citep{BM04}.
Another possible extension of the present approach is to include the
more general case of polymerization or polycondensation reactions of
nucleotides \citep{San03,BAHN} and of peptides \citep{Plasson,BLL07}.
The polycondensation reactions can easily be included and would simply
increase the number of possible reactions from seven in the present
work to any arbitrary number.
Particularly useful would be the study of network catalysis
\citep{PB10,Hoc17}, which may be strongly affected by fluctuations having
either an enhancing or diminishing effect on achieving homochirality.

\section*{Acknowledgements}

The author thanks Hannah Zona for comments and encouragement.
I am indebted to the two reviewers for their constructive remarks
and suggestions.
This work was supported through
the National Science Foundation, grant AAG-1615100,
the University of Colorado through its support of the
George Ellery Hale visiting faculty appointment,
and the grant ``Bottlenecks for particle growth in turbulent aerosols''
from the Knut and Alice Wallenberg Foundation, Dnr.\ KAW 2014.0048.
The simulations were performed using resources provided by
the Swedish National Infrastructure for Computing (SNIC)
at the Royal Institute of Technology in Stockholm and
Chalmers Centre for Computational Science and Engineering (C3SE).


\end{document}